\documentclass{article}

\usepackage{PRIMEarxiv}

\usepackage[utf8]{inputenc} 
\usepackage[T1]{fontenc}    
\usepackage{hyperref}       
\usepackage{url}            
\usepackage{booktabs}       
\usepackage{amsfonts}       
\usepackage{nicefrac}       
\usepackage{microtype}      
\usepackage{lipsum}
\usepackage{fancyhdr}       
\usepackage{graphicx}       
\graphicspath{{media/}}     

\makeatletter
\renewcommand{\subsubsubsection}[1]{\paragraph{#1}\mbox{}\\}
\everymath{\displaystyle}
\usepackage{float}
\newtheorem{definition}{Definition}

\usepackage{mathtools}
\usepackage[english,lined,boxed,ruled]{algorithm2e}

\title{Crosswell seismic tomographic inversion using NSGA II
}

\author{
  Arthur Anthony da Cunha Romão E Silva \\
  Faculdade de Ciências da Saúde do Trairi (FACISA) \\
  Universidade Federal do Rio Grande do Norte \\
  Santa Cruz-RN\\
  \texttt{arthur.romao@ufrn.br} \\
   \And
  Francisco Márcio Barboza \\
  Centro de Ensino Superior do Seridó (CERES) \\
   Universidade Federal do Rio Grande do Norte\\
  Caicó - RN\\
  \texttt{marcio.barboza@ufrn.br} \\
}

\begin{document}
\maketitle

\begin{abstract}
In the present paper, the solution of the seismic data inversion problem through multi-objective optimization with NSGA II is addressed. The seismic inversion consists of estimating the slowness of rocks in the subsurface from the travel times of the waves through the transmitters and receivers of the seismic waves. The inversion process uses the multi-objective optimization technique, in which it is necessary to estimate the trade-off between the misfit functional and the stabilizing constraint. The constraint used was the classic first-order Tikhonov. A synthetic model of inclined layers was chosen. The results obtained were acceptable. Based on the graph of the cumulative errors related to the solution, it was estimated correctly, that is, the optimization method used was effective to obtain acceptable results, in addition to the constraint solving the instability problem of the inverse problem treated. Another important result was the achievement of a convergence curve on the Pareto frontier, an important result for future research.
\end{abstract}


\section{Introduction}
Inverse geophysical problems can be put in the form of a single-objective optimization problem, which is premised on finding a solution in a search space through an objective function. These are solved through the weights (or weighted sum) in which all objective functions are linearly combined into a single functional. This consists of two parts, the misfit, how much closer the estimated solution is to the true one, and a geological constraint on priori, which in this case is the constraint. In this case, we use a weight for the constraint, called the Lagrange multiplier $\lambda$, which is an indication factor or how much of the constraint will be incorporated into a solution, chosen through the method of curvature L, being normally used in linear problems \cite{lawson1995solving}. However, in this method, the inversion process is complex and requires a priori choice of several parameters. This mode is indicated for vectorized multi-objective optimization, which eliminates the need to choose these weights.

Some papers using multi-objective optimization techniques can be cited \cite{fonseca1993genetic, schaffer1985multiple, justesen2009multi, zitzler2000comparison}. This optimization method has been shown to be effective in several engineering problems, computing and geophysics problems of joint inversion \cite{kozlovskaya2007joint, dal2007joint, deb2000fast, schaffer1985multiple}. However, there are not so many published papers with these algorithms with tomographic inversion and constraint, making a more in-depth study of this method necessary.

The vectorized multi-objective optimization is modeled from a function that takes as a parameter two separate objectives subject to a vector function of constraints. These are often in conflict with each other. A mono-objective inverse geophysical problem can be transformed into a multi-objective one, for that we transform the constraint into an objective, in this way we have two objectives, that of finding an acceptable proximate solution, and that of incorporating the constraint to the solution \cite{cohon2004multiobjective}.

It is based on the discovery of solutions that satisfy a trade-off between the misfit functional and the application of the constraint. The two-dimensional Pareto front is created, which is the solution of a multi-objective algorithm, in which it has the acceptable optimal solutions, that is, the non-dominated solutions \cite{miettinen2012nonlinear}. The Pareto front remains relatively unexplored in both linear and nonlinear problems. 

Bioinspired global optimization methods are stochastic techniques to obtain minimization or maximization solutions through mathematical mechanisms extracted from nature. Like genetic algorithms, which are based on the genetic evolution of individuals \cite{gaspar2008robustness}. Swarm Intelligence Algorithms, which are based on the personal knowledge of individuals and cultural transmission in the pursuit of an objective \cite{coello2002mopso}. And hunting-based algorithms, based on a hierarchy and collaboration of a group of animals to capture the best prey \cite{mirjalili2016multi}.

The objective of this paper is to solve the tomographic inversion problem using the multi-objective genetic algorithm, with the use of straight ray modeling and the application of first order Tikhonov regularization. 

\section{Multi-objective Optimization}

The process of systematically and simultaneously optimizing a collection of objective functions is called multi-objective optimization (MO). The inversion using a misfit and a constraint can be seen as a multi-objective optimization .

The MO can be defined as the minimization of a vector objective function $F(\textbf{s}) = [F_1(\textbf{s}), F_2(\textbf{s}), ..., F_k(\textbf{s})]$ which takes values in objective space $\mathbb{R}^k$ subject to $\textbf{s} = (s_1, s_2, ..., s_n) \in S \subset \mathbb{R}^n$, where $S$ is a set of feasible solutions (parameter search space) of the multi-objective optimization problem.

\begin{equation}
    \min_{\textbf{s} \in S} F(\textbf{s}) = [F_1(\textbf{s}), F_2(\textbf{s}), ..., F_k(\textbf{s})]
    \label{eqn_1}
\end{equation}

$k$ is the number of objective functions, $s \in S$ is a vector of parameters (also called decision variables), where $n$ is the number of independent variables $s_i$. $F(\textbf{s}) \in \mathbb{R}^k$ is a vector of objective functions $F_i(\textbf{s}) : \mathbb{R}^n \rightarrow \mathbb{R}$ . $F_i(\textbf{s})$ are also called objective functions or cost functions.

In the case of tomographic seismic inversion, the idea is to use simultaneously the optimality of the misfit, that is, the value of the estimated solution, and the incorporation of the constraint, tracing a Pareto frontier on $\mathbb{R}^2$, thus selecting empirically from a set of iterations an acceptable solution. We can for example incorporate more than one constraint in this vector model, in this case we will obtain a Pareto surface in $\mathbb{R}^3$. Since no optimal solution, denoted by $s^*$, can in general simultaneously minimize each $F_i$ , an optimization concept that is useful in the multi-objective context is Pareto optimality.

\subsection{Pareto Front}

In multi-objective optimization it is generally impossible to satisfy all objectives simultaneously. This means that for the multi-objective problem \ref{eqn_1}, it is highly unlikely to have a single $s*$ that minimizes all $F_i$ simultaneously; therefore, the solution is defined in terms of Pareto optimality in the following sense \cite{cohon2004multiobjective}: a feasible solution to a multi-objective optimization problem is Pareto optimal (non-inferior, non-dominated) if no other solution exists feasible that will produce an improvement in one objective without causing a degradation, at least in another objective. More formally we have the following definitions \cite{miettinen2012nonlinear}:

\begin{definition} 
(Pareto optimality) A decision vector $ \textbf{s*} \in S $ is a Pareto optimal (efficient, non-inferior, non-dominated) if there is no other decision vector $ \textbf{s} \in S $ such that $F_i(\textbf{s}) \leq F_i(\textbf{s*})$ for all $i = 1, ... , k$ and at least one of the inequalities is strict. An objective vector $ F(\textbf{s*}) = \hat{z}$ is Pareto optimal if the corresponding decision vector $ \textbf{s*} \in S $ is Pareto optimal.
\end{definition}

The Pareto optimality concept can be used to determine decision vectors that can be mathematically considered as potential solutions to the problem described in the equation \ref{eqn_1}. 

\begin{figure}[H]
\centering
\includegraphics[width=0.6\linewidth]{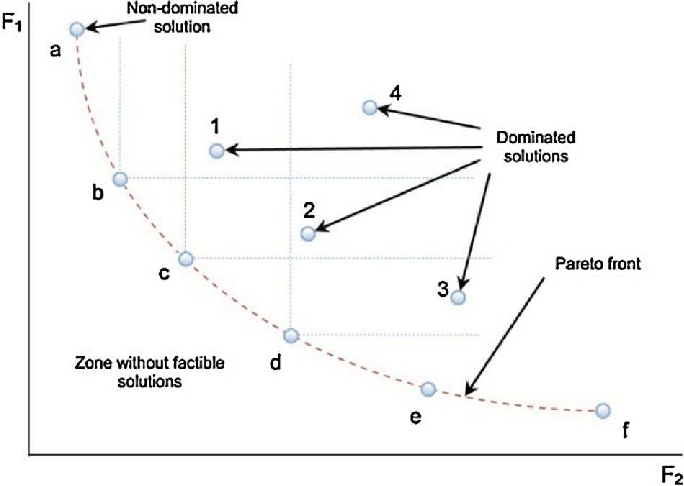}
\caption{Dominated and non-dominated solutions}\label{fig:otimalidade_fronteira_pareto}
\vspace{0.3cm}\textbf{Source}: \cite{asadi2019optimisation}
\end{figure}

The Fig. \ref{fig:otimalidade_fronteira_pareto} illustrates the concept of dominance and Pareto optimal set, that is, a set formed by Pareto optimal solutions (Pareto frontier). 

In applications, the decision maker is often presented with several Pareto optimal points, from which the one that offers the best trade-off between the multiple objectives is selected. Theoretically, the more Pareto optimal points available (ideally the entire Pareto set), the better the final choice.

The set of Pareto optimal solutions of a multi-objective optimization problem is called the Pareto frontier. Depending on the problem, we can see such a set graphically, in objective space. The objective space differs from the parameter space, which is defined by the decision variables of the problem, as follows: the axes are represented by the objective functions of the problem, so each point in this space has its coordinates defined by the values of the objective functions evaluated in a point in the parameter space. 

We often want to get the entire Pareto front, or at least an approximation of it. Although there are several optimal solutions found in the problem, we know that only one will be chosen in the end. This chosen solution is called the best compromise solution. The ratio of the amount that must be increased for one objective in order to decrease another objective is called the trade-off. Trade-offs and Pareto optimal solutions are two types of important information for the decision maker to choose the best compromise solution. 

\subsection{Global Optimization Methods}

In this paper the crosswell seismic inversion is posed as a vectorized multi-objective problem. To generate Pareto frontier we opted for linear global optimization methods. We can enumerate some advantages of this methodology: 

\begin{itemize}
    \item Ease of implementation;
    \item They are suitable for features in non-convex problems;
    \item Do not need to choose weights a priori;
    \item Flexibility in the mathematical formulation of the problem. 
\end{itemize}

Genetic Algorithm (GA) is a method whose mathematical evolution aims to reproduce or imitate the evolutions of complex evolutionary optimization systems existing in nature; in other words, it is based on analogies with the behavior of complex natural systems, in this case the evolutionary system proposed by Charles Darwin \cite{darwin1909origin}. In the present paper, we used it for the GA in its multi-objective version Non-dominated Sorting Genetic Algorithm II (NSGA II). 

\subsubsection{Non-dominated Sorting Genetic Algorithm II}

Introduced by Holland \cite{holland1992adaptation} and Goldberg \cite{golberg1989genetic}, GA is a global optimization algorithm based on genetic evolution and natural selection. These processes are premised on the following aspects: replication, mutation and competition. A fitter individual (solution) has a greater chance of surviving and passing on its gene, while the reverse is also true. These are interpreted from genes combined with chromosomes. The genetic operators basically consist of the crossing between them and the application of the mutation on a portion of the generated children. The fittest individuals are selected to participate in the next generation. Some strategies such as elitism bring together a part of the best parents of the previous generation to participate in the new solution, in order to avoid discarding promising solutions.

\begin{figure}[H]
\centering
\includegraphics[width=0.6\linewidth]{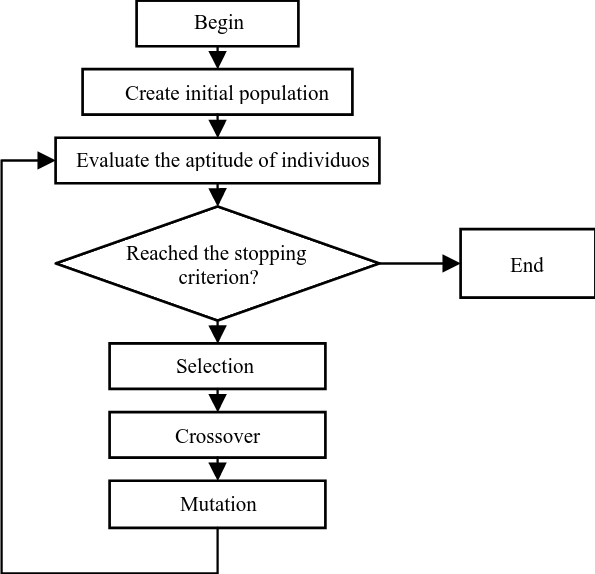}
\caption{Genetic Algorithm Flowchart}\label{fig:fluxograma_ga}
\end{figure}

The NSGA-II method is a multi-objective GA, presented by \cite{deb2002fast}, which classifies solutions according to the Pareto dominance concept. In this way, using elements of GA's such as the succession of generations and population structure, at the end of the optimization process a set of non-dominated solutions is obtained to be submitted to decision making by the interpreter. The main mechanisms introduced in the NSGA-II algorithm are: 

\begin{itemize}
    \item Elitism, which guarantees the preservation of good solutions in the search process;
    \item The Fast Non Dominated Sorting (FNS) procedure in which the population is classified into different levels according to Pareto dominance;
    \item The Crowding Distance Assignment (CDA) procedure that aims to ensure population diversity.
\end{itemize}

The general functioning of the NSGA-II method consists of generating an initial random population $P_i$ of dimension $p_s$ and, through the genetic operators, a population of descendants $Q_i$ also of dimension $p_s$. These populations $P_i$ and $Q_i$ are then combined, forming a set $R_i$ with $2xp_s$ individuals. The expanded population is classified according to dominance by the FNS procedure and the next population $P_{i+1}$ is already composed of the elements of $R_i$ with the lowest dominance index, starting with the non-dominated individuals of the first Pareto frontier associated with these elements and so on. If the last frontier to have its individuals added to the population $P_{i+1}$ has a number of solutions that, if added to those already added, exceeds the size $p_s$ of $P_{i+1}$ the surplus individuals are discarded according to the diversity criterion, that is, those with the highest value of the distance index are chosen to compose $P_{i+1}$. Finally, after defining the new population $P_{i+1}$ of dimension $p_s$, a population of descendants $Q_{i+1}$ is created using the genetic operators of selection, recombination and mutation. The process is then repeated until the stopping criterion is satisfied.

\begingroup
\begin{algorithm}[H]
		\SetAlgoLined
		Generate population randomly\;
		Generate a set of non-dominated solutions\;
		Evaluate fitness of each particle contained in the population\; 
		\While{Stop criterion is not satisfied }{
		    Generate population of children by crossing\;
		    Apply mutation on children\;
		    Unite parents and children into a new population\;
			\ForEach{individual $\textbf{s}_{i}$ in the new population}{
				Assign rank to each solution using Non Dominated Sorting\; 
			}
			Create non-dominated solutions from the population\; 
			Execute Crowding Distance\; 
			Include best solutions in the non-dominated solution set\;
			}
		return the best set of non-dominated solutions\;
		\caption{Non-dominated Sorting Genetic Algorithm II}
		\label{alg:nsga2}
	\end{algorithm}
	\endgroup

The preservation of density in the original NSGA was tied to the use of the share function approach, however, there are two difficulties in this approach which are: the proper choice of the value of the share parameter and the total complexity of this approach. NSGA-II replaces the share function approach with a compression comparison approach that eliminates the above difficulties as it does not require any user-defined parameters to maintain diversity among population members and improves computational complexity. 

\subsubsubsection{Structure of a Chromosome}

\begin{figure}[H]
\centering
\includegraphics[width=0.6\linewidth]{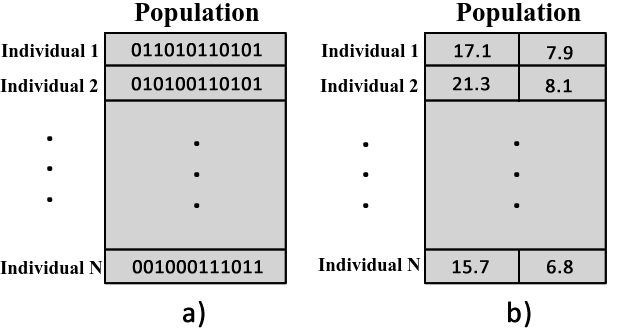}
\caption{Genetic Structure of an Individual }\label{fig:estrutura_cromossomo}
\vspace{0.3cm}\textbf{Source}: Adapted from \cite{grando2018classificaccao}
\end{figure}

An individual is interpreted as a chromosome, which is vectorially defined as a set of combined genes. Defined from $\mathbb{R}^n$, $\textbf{s}_i = [\textbf{s}_i1, \textbf{s}_i2, ... \textbf{s}_n]$. A set of these individuals composes a population, and at each iteration of the GA a new generation is created. The Fig. \ref{fig:estrutura_cromossomo} denotes two representations that chromosomes can be encoded in a population, solutions that are discretized into binary numbers, and solutions that are encoded using real numbers.

\subsubsubsection{Selection}

In the selection process, based on genetic evolution, individuals with better survival aptitude tend to pass their gene on, and even be used in the next generation (elitism). The fitness of an individual is calculated through a function, and it will map how much individuals are moving towards the solution. There are several methodologies for selecting the best individuals, of which roulette and the tournament can be cited \cite{faceli2011inteligencia}.

In tournament selection a set of individuals are chosen at random for the dispute, the fittest are selected to be reproduced, some algorithms include elitism, capable of storing a portion of the best individuals for the next generation \cite{linden2005algoritmos}.

\begin{figure}[H]
\centering
\includegraphics[width=0.6\linewidth]{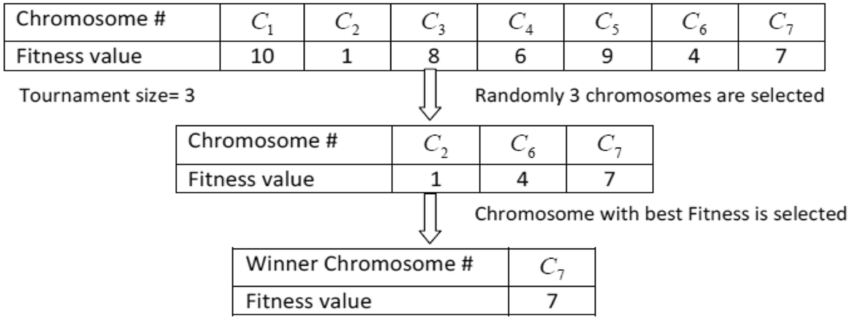}
\caption{Selection by Tournament}\label{fig:torneio}
\vspace{0.3cm}\textbf{Source}: \cite{banihashemian2021novel}
\end{figure}

\subsubsubsection{Crossover}

\begin{figure}[H]
\centering
\includegraphics[width=0.6\linewidth]{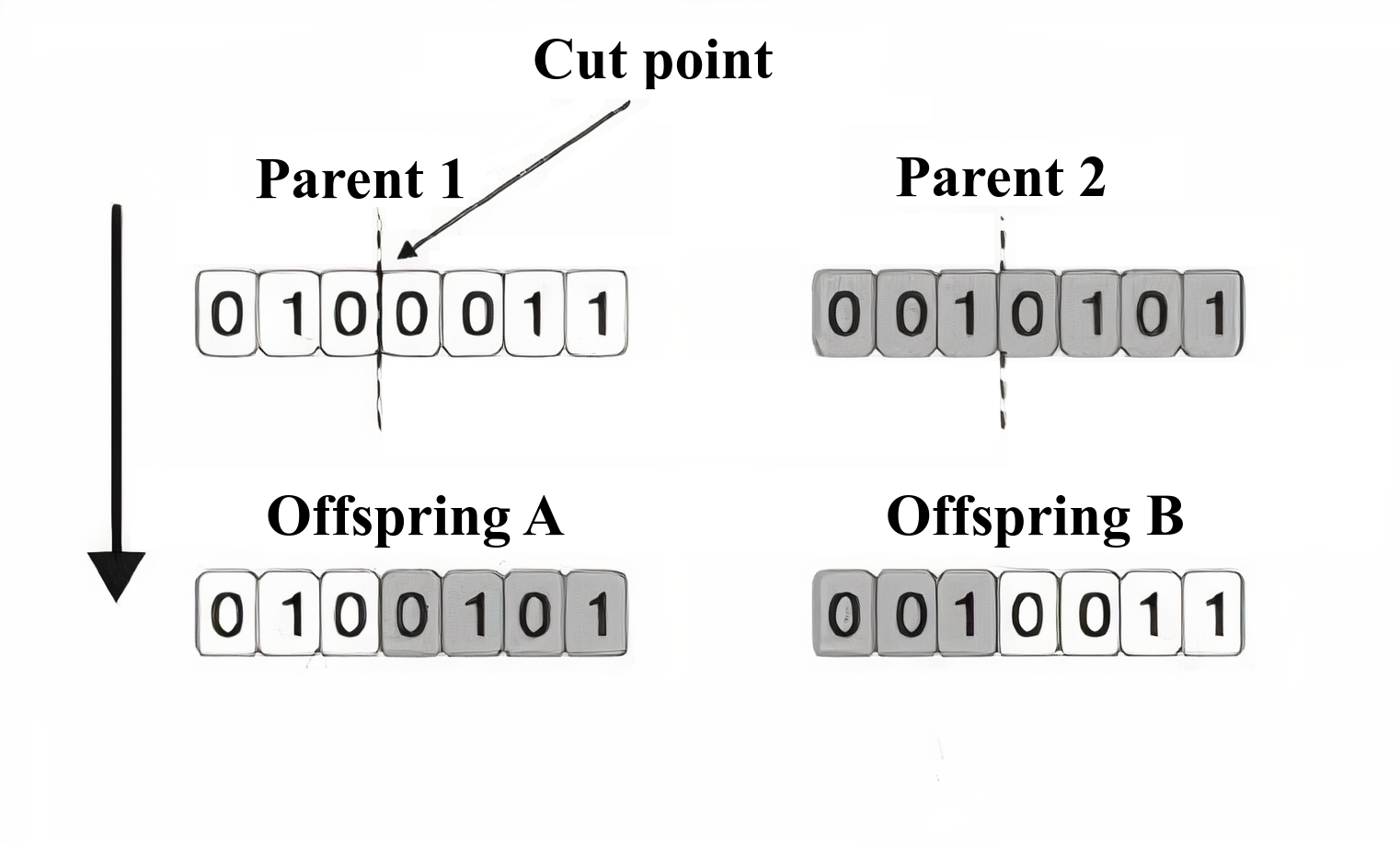}
\caption{Crossover Genetic Operator}\label{fig:cruzamento}
\vspace{0.3cm}\textbf{Source}: Adapted from \cite{faceli2011inteligencia}
\end{figure}

The genetic crossover operation consists of randomly selecting two chromosomes (individuals) from a population, and performing the genetic exchange at random cut-off points of each individual, fragmenting them into two segments, thus joining the combination of genes from these chromosome segments for the generation of two new children, as shown in the Fig. \ref{fig:cruzamento}.

\subsubsubsection{Mutation}

\begin{figure}[H]
\centering
\includegraphics[width=0.6\linewidth]{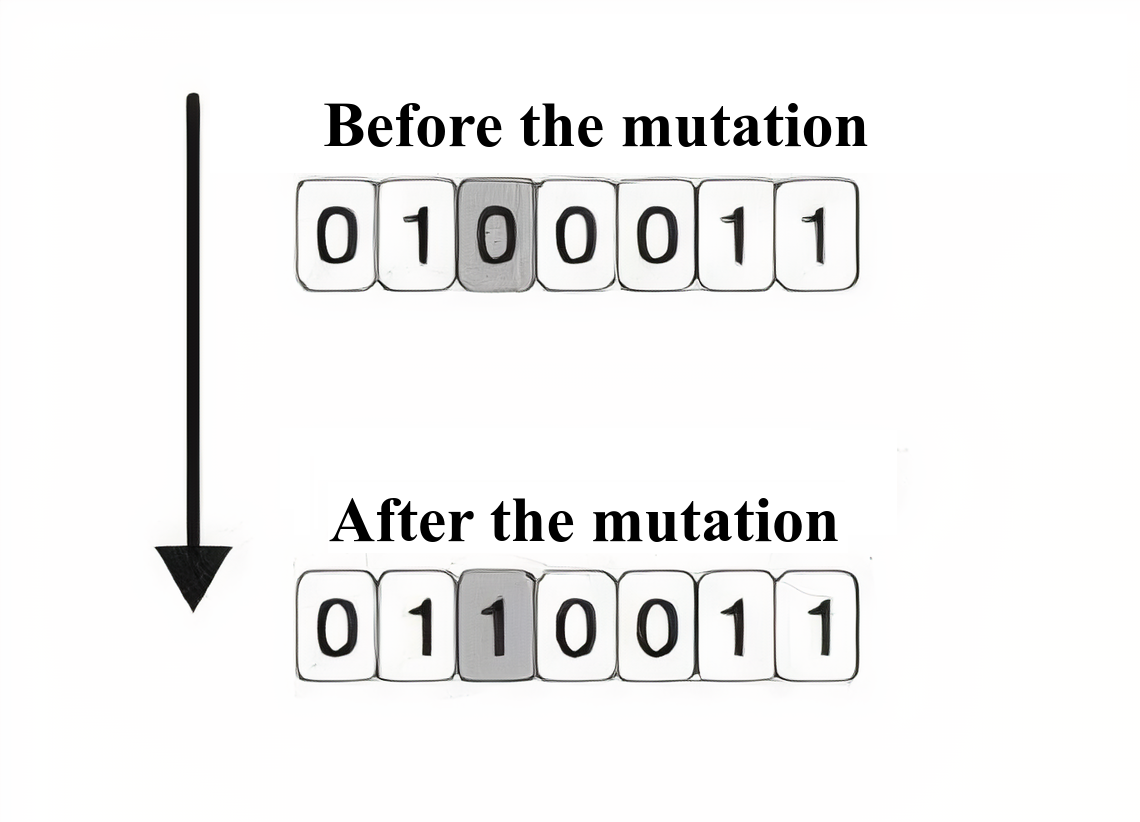}
\caption{Genetic Mutation Operator}\label{fig:mutacao}
\vspace{0.3cm}\textbf{Source}: Adapted from \cite{faceli2011inteligencia}
\end{figure}

Mutation (see Fig. \ref{fig:mutacao}) consists of a technique to mutate a segment of genes from a portion of chromosomes in a population. This procedure guarantees the genetic variability of individuals, allowing to aggregate and innovate promising solutions to the optimization process. This process occurs after crossing over, and the factor that will define which individuals will be mutated is the mutation rate.

\subsubsubsection{Ranking and Non Dominated Sort}

\begin{figure}[H]
\centering
\includegraphics[width=0.6\linewidth]{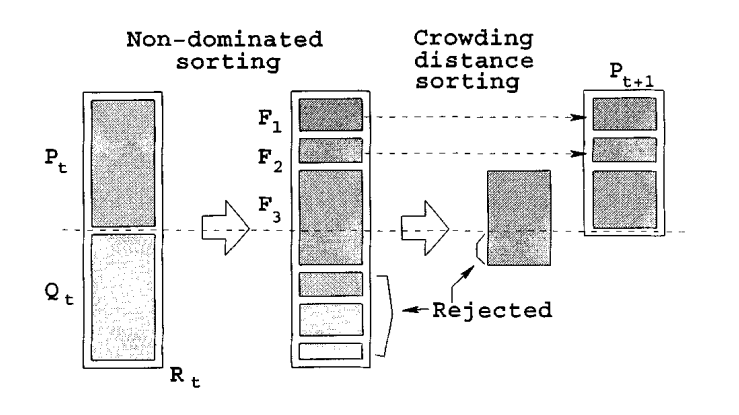}
\caption{Non Dominated Sort Process for Ranking}\label{fig:non_dominated_sorting}
\vspace{0.3cm}\textbf{Source}: \cite{deb2002fast}
\end{figure}

The Non Dominated Sort procedure consists of comparing and classifying a set of solutions based on the dominance concept. According to the Fig. \ref{fig:non_dominated_sorting}, initially a set of solutions (first generation) $P_0$ is generated, and after the genetic operations we have the descendants $Q_0$, these are filtered and incorporated into the Pareto Frontier through the algorithm, ranking the dominance level of each solution. 

This protocol occurs successively in the other generations, in which these Frontier solutions are replaced through aptitude classification. Each generated frontier is enumerated with a rank. Individuals are selected from a comparison called crowded, in which each compares the solutions pair by pair using the rank of these solutions as a criterion. And then the procedure of Crowding Distance is performed. Selecting the most apt, and rejecting the least promising solutions. 

\subsubsubsection{Crowding Distance}

\begin{figure}[H]
\centering
\includegraphics[width=0.6\linewidth]{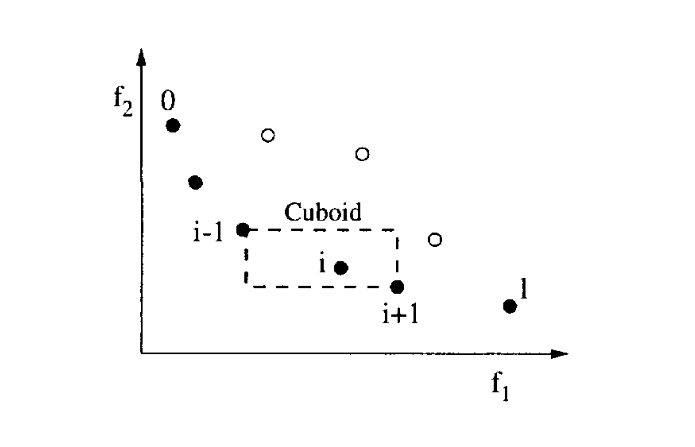}
\caption{Crowding Distance Technique}\label{fig:crowding_distance}
\vspace{0.3cm}\textbf{Source}: \cite{deb2000fast}
\end{figure}

During the NSGA-II heuristic process, a set of solutions is classified through an ordinate ranking. The Crowding Distance is a technique incorporated to this one that guarantees a metric sparity between the non-dominated solutions in the Pareto Frontier. In this way it becomes more widespread, diversifying the solutions, this is necessary because the solutions often remain concentrated in certain points of the Border \cite{narino2014otimizaccao, marinho2009aplicaccao}. As shown in the Fig. \ref{fig:crowding_distance}, solutions have a cuboid that delimits their ends.

\section{Crosswell Seismic Tomography}

Crosswell seismic tomography is one of several techniques for inferring geological data through inversion of existing seismic data within the Earth. Commonly used when it comes to imaging exploration of geological structures, exploration of mineral resources and archaeological sites \cite{metwaly2005combined, gustavsson1986seismic, ajo2009optimal, ajo2007applying, byun2010crosswell}. The first papers on seismic tomography took place through the contributions proposed by \cite{backus1970uniqueness} in the year 1970. Some were important for the development of the technique and a study of possible limitations \cite{white1989two, lanz1998refraction, menke1984resolving, goudswaard1998detection}. It has also been used for the study of radioactive waste disposal areas \cite{peterson1985applications}, $CO_2$ injection projects \cite{ajo2009optimal,byun2010crosswell}, mineral exploration in \cite{gustavsson1986seismic} mines, among others. 

The objective is to obtain models of rock slowness by imaging geological structures of a given region of the subsurface using waveforms (either mechanical or electromagnetic) or travel times \cite{fernandez2012reservoir,tronicke2012crosshole,ebrahimi2018nonlinear}. 

\begin{figure}[!htbp]
\centering
\includegraphics[width=0.6\linewidth]{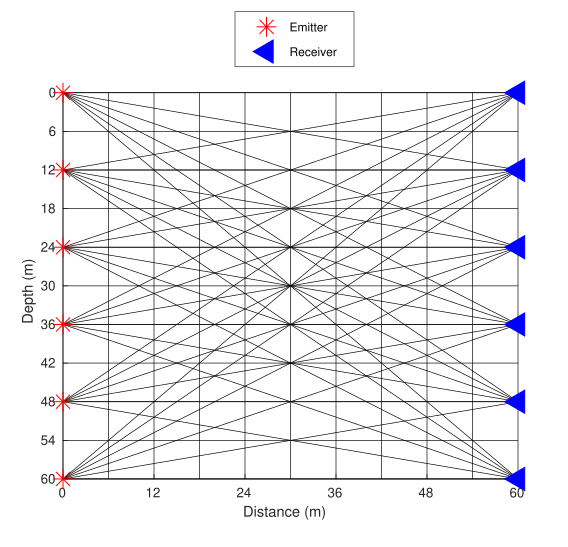}
\caption{Geometric simulation of acoustic waves passing through the subsurface from emitters to receivers}\label{fig:wave_sample_tomography_sismic}
\vspace{0.3cm}\textbf{Source}: \cite{arthur2021tomography}
\end{figure}

The observed seismic data is measured by drilling adjacent wells, where the time in which the signals travel from each emitter to each receiver is obtained. It is possible to obtain heterogeneous models, with discontinuities in slowness, or homogeneous models. The computational modeling of the path in which the signals (seismic waves) cross the subsurface adopted in this paper was that of straight rays. These follow a discretized path in cells, called pixels (see Fig. \ref{fig:wave_sample_tomography_sismic}). In this way we obtain the distances between the signal segments in the mesh \cite{briao2005tomografia}.

Each seismic wave suffers a certain perturbation in the discretized grid depending on the geological properties of a given region in the subsurface. Then the direct problem is to obtain the vector of transit times from the knowledge of the slowness. The equation \ref{eqn_2} expresses how these times are obtained.

\begin{equation}
 G(\textbf{s}) = t
 \label{eqn_2}
\end{equation} 

Where $G$ is the straight ray modeling operator determined by the discretization of the problem, $\textbf{s}$ is the vector of parameters (model) of size $n$, and $t$ is the vector of transit times observed.

\begin{equation}
    \min_{\textbf{s} \in S} F(\textbf{s}) = [\Phi(\textbf{s}), \mathcal{R}(\textbf{s})]
    \label{eqn_3}
\end{equation}

\begin{equation}
  \Phi =\frac{\| G(\textbf{s})-t_{obs} \|_2 ^2}{N_{obs}}
  \label{eqn_4}
\end{equation}

\begin{equation}
  \mathcal{R} = \|D\textbf{s}\|_2^2  \label{eqn_5}
\end{equation}

The equation \ref{eqn_3} expresses the minimization of the vector function that contains two objectives, the first objective measures the misfit of the data, which is the $\Phi$ (expressed in the equation \ref{eqn_4}), where $t_{obs}$ the observed data and $N_{obs}$ the number of these observations. In the second objective (expressed in the equation \ref{eqn_5}) it measures the smoothness of the model, denoted by the symbol $\mathcal{R}$, in which $D$ is a finite difference matrix. 

The search for the geological properties of a given region is found from this modeling of a minimization problem. In which we intend to estimate the slowness vector that explains the observed seismic data. This is done from a misfit function, in which the obtained model theoretically approaches the true model is compared. 

Mathematically, many parameter vectors can adjust this criterion, in addition to considering that the seismic data may have undesirable noise incorporated, and this makes the problem ill-posed, being in these situations necessary the application of a regularization methodology. In this case, we incorporate the optimization mathematics into the constraint factor, which in this paper we use classic first-order Tikhonov's. 

Usually in a scalarized approach to the optimization problem, a balancing factor, called Lagrange multiplier, is applied to it, where the more it is applied, the more the solution will obtain the bias of the chosen constraint, and the smaller, the more unstable the solution. But in a multi-objective approach, it eliminates the need to apply this criterion, transforming the incorporation of a constraint as an objective, solving the problem by generating a Pareto frontier with solutions that minimize the misfit function and the constraint. 

\section{Results and discussions}
In this section we present information about the parameters used in the seismic data inversion process, the tomograms, the Pareto frontier and a statistical analysis of the cumulative relative errors. A synthetic geological model was chosen to carry out the experiment. This one had as a strategy the application of global optimization with the multi-objective algorithm NSGA II and the application of a regularization method. The observed data were generated from the modeling of straight rays, and we applied to these $5 \%$ of pseudorandom Gaussian noise. 

The chosen model was discretized in $100$ parameters, the imaging consists of a matrix of order $10$ where each index $ij$ consists of the rock velocities. The search for the estimated parameters of the models during the inversion of the seismic data was carried out by the interval of delays of $50\%$ more and $50\%$ less than the parameters of the true models. The parameters of the NSGA II algorithm are detailed in table I.

\begin{table}[H]
\centering
\bgroup
\def\arraystretch{1.5}
\begin{tabular}{|c|l|c|l|c|l|}
\hline
\multicolumn{2}{|c|}{\textbf{$P_c$}} & \multicolumn{2}{c|}{\textbf{$P_m$}} & \multicolumn{2}{c|}{\textbf{$\sigma$}} \\ \hline
\multicolumn{2}{|c|}{$1$}            & \multicolumn{2}{c|}{$0.1$}          & \multicolumn{2}{c|}{$0.1$}             \\ \hline
\end{tabular}\label{tab:tbl_params_ga}\egroup
\vspace{0.5cm}
\caption{GA control parameters}
\end{table}
The factor \textbf{$p_s$} is related to the population, and is the most significant, as it will provide a genetic variety (characters). The \textbf{$k$} refers to the maximum number of generations. The crossover ($P_c$) and mutation rate ($P_m$) and the sigma ($\sigma$) are important to generate innovation on the part of genes, there are many crossover strategies, such as elitism, that adds part of the parents in the new generation. These inversion coefficients are detailed in table 2.

\begin{table}[H]
\centering
\bgroup
\def\arraystretch{1.5}
\begin{tabular}{|c|l|c|l|c|l|c|l|}
\hline
\multicolumn{2}{|c|}{\textbf{$p_s$}} & \multicolumn{2}{c|}{\textbf{$k$}} & \multicolumn{2}{c|}{\textbf{$n$}}  \\ \hline
\multicolumn{2}{|c|}{$1000$}         & \multicolumn{2}{c|}{$1000$}              & \multicolumn{2}{c|}{$100$}         \\ \hline
\end{tabular}\label{tab:tbl_params}\egroup
\vspace{0.5cm}
\caption{Inversion Coefficients}
\end{table}

\subsection{Inclined Layers Model}
\begin{figure}[!htbp]
\centering
\includegraphics[width=0.6\linewidth]{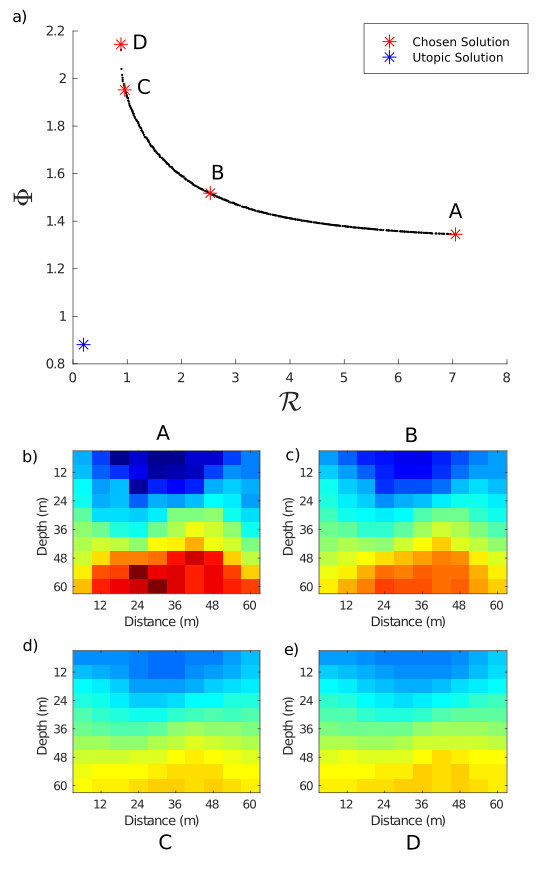}
\caption{Inversions result}\label{fig:modelo_2_inversions}
\end{figure}

The model is based on \cite{dantas2015resoluccao}, characterized by having a geological structure in inclined layers, without failures or abrupt changes, but with a smooth consistency in the parameter velocities. A discussion was made about the solutions obtained in three regions of the Pareto frontier, as well as a statistical analysis of the solutions from the relative cumulative errors.
\begin{figure}[H]
\centering
\includegraphics[width=0.6\linewidth]{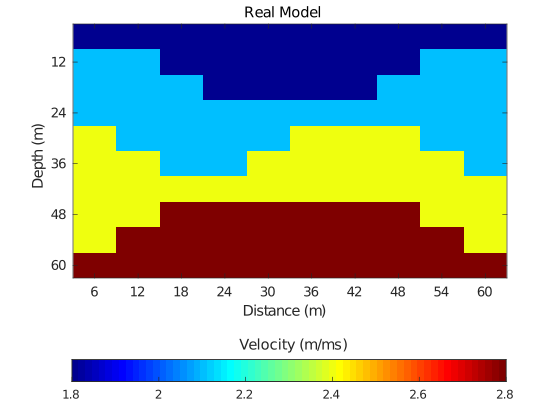}
\caption{Inclined Layers Model}\label{fig:modelo_2}
\end{figure}
The Fig. \ref{fig:modelo_2_inversions} shows the results obtained from the inversions used in the model in the four regions of the Pareto frontier. It is visible that there are regions that converge to a more adequate solution based on the trade-off between the minimization and the incorporation of the constraint. Based on this, we conclude that the Smoothness constraint proved to be effective in this context to regularize in the four situations, in effect making the problem well posed. Thus, satisfactory results were obtained when estimating the true model.

\begin{figure}[H]
\centering
\includegraphics[width=0.6\linewidth]{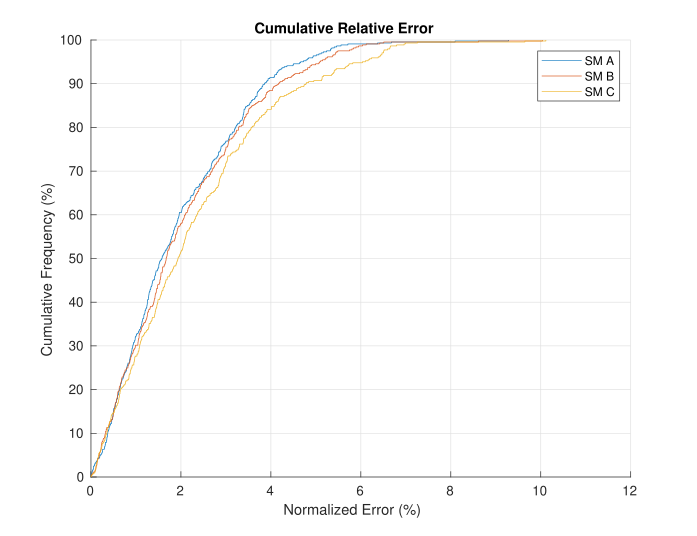}
\caption{Relative cumulative error figures}\label{fig:4_figures_article_model_2}
\end{figure}

The Fig. \ref{fig:4_figures_article_model_2} shows the relative cumulative error, it is noticeable that the inversion of the model for the three feasible points obtained an accuracy of 6\% of the cumulative error in more than 94\% of the data, this shows that the function objective was minimized and the constraint was incorporated into the solution in this region of the Pareto frontier, being it stable.
\section{Conclusion}
The smooth model by inclined layers was used as a sample of the synthetic parameters. Matlab software was used to produce the inversion algorithms, straight ray modeling, as well as the NSGA II global multi-objective optimization algorithm. The results obtained were the distribution of velocities through the tomograms, an evaluation of the Pareto frontier and a statistical analysis of the relative cumulative errors to verify whether optimal solutions were obtained, that is, closer to the true parameters. 

The results showed better convergence in a region on the Pareto frontier, an important fact for future research. The conclusion came from the statistical analysis of the relative cumulative errors of the inversions performed.

\bibliographystyle{unsrt}  
\bibliography{templateArxiv}

\end{document}